\documentclass[aps,twocolumn,showpacs,floatfix]{revtex4} 

\usepackage{graphicx}

\begin{document}

\title{Interface relaxation in electrophoretic deposition of polymer
chains: Effects of segmental dynamics, molecular weight, and 
field}

\author{Frank W. Bentrem, Jun Xie, and R.~B. Pandey}

\affiliation{Department of Physics and Astronomy, The University of Southern Mississippi,
Hattiesburg, MS 39406-5046}

\date{\today}

\begin{abstract}

Using different segmental dynamics and relaxation, characteristics of the
interface growth is examined in an electrophoretic deposition of polymer
chains on a three (2+1) dimensional discrete lattice with a Monte Carlo
simulation.  Incorporation of faster modes such as crankshaft and
reptation movements along with the relatively slow kink-jump dynamics
seems crucial in relaxing the interface width. As the continuously
released polymer chains are driven (via segmental movements) and
deposited, the interface width $W$ grows with the number of time steps
$t$, $W \propto t^{\beta},$ ($\beta \sim 0.4$--$0.8)$, which is followed 
by
its saturation to a steady-state value $W_s$.  Stopping the release of
additional chains after saturation while continuing the segmental
movements relaxes the saturated width to an equilibrium value ($W_s \to
W_r$). Scaling of the relaxed interface width $W_r$ with the driving field
$E$, $W_r \propto E^{-1/2}$ remains similar to that of the steady-state
$W_s$ width. In contrast to monotonic increase of the steady-state width
$W_s$, the relaxed interface width $W_r$ is found to decay (possibly as a
stretched exponential)  with the molecular weight.

\end{abstract}


\maketitle

\section{Introduction}

The deposition process \cite{barabasi, family} is one of the common 
methods used for growing and designing composites, polymeric materials, 
interfaces, and surface coating \cite{wool}.
As the polymer chains are driven toward an impenetrable
substrate/wall, the polymer density at the substrate grows and the 
interface
develops \cite{bpf, fp1, fp2}. A number of parameters (e.g., temperature, 
pressure/field, molecular weight of polymer) and processes (e.g., rate 
of 
polymer release, segmental dynamics, relaxation, etc.) affect the growth 
of the
polymer density and its interface. 
Recently we examined the growth of the interface, its scaling,
and roughness in an electrophoretic deposition model on a discrete 
lattice
using kink-jump segmental dynamics \cite{bpf} and by including
reptation \cite{fp1, fp2}. 
We have observed many interesting scaling behaviors of the interface 
width
in these continuous deposition processes where the polymer chains
are released at a constant rate throughout the computer simulation.

As the polymer chains deposit, the polymer density spreads from the 
substrate
toward the source of the polymer chains, and the interface width 
grows.
Even in such a continuous deposition process, the interface width $W$ 
reaches a steady-state (saturated) value ($W_s$) after an appropriate 
growth period. The steady-state width $W_s$ exhibits interesting scaling 
with 
temperature, field, and molecular weight leading to
roughening and deroughening \cite{bpf, fp1, fp2}.

In this article we examine the characteristics of the density profile 
and
the interface width as we relax the system (which includes the polymer
bulk and interface). The study of relaxations at polymer
surfaces is relatively new \cite{liu97,wallace01}, and characteristic
surface relaxation times $\tau$ have been measured for some polymer materials
\cite{riemann96,riemann97,tanaka00}. One approach to achieve such a relaxation
in simulations is to stop injecting new polymer chains after an appropriate amount of
polymer chains are in the system but continue to allow segmental motion of
the chains.
Temperature $T$, field $E$, and molecular weight $L_c$ play 
important
roles in relaxing the interface width; here we restrict the analysis to 
the effects of 
field
and molecular weight. 
We also probe the relaxation by examining the effects of incorporating 
the
faster modes of movements such as crankshaft and reptation along with
the 
relatively slow kink-jump segmental dynamics \cite{binder}. In the next section 
we 
briefly describe the model followed by results and discussion, and 
finally, a summary/conclusion.  

\section{Model}

We briefly describe the model to point out the differences between the 
simulation procedure adopted here for segmental dynamics and 
relaxation
and our previous studies \cite{bpf, fp1, fp2}. We consider a discrete 
lattice of size
$L_x \times L \times L$ with a large aspect ratio $L_x/L$. Typically,
$L_x = 100$--$200$, $L = 40$, $60$. Polymer chains each of length $L_c$, 
generated
as the trail of a random walk of $L_c$ steps along the lattice with 
excluded 
volume constraints, are released from the $x=1$ end of the sample. The 
lengths used here ($L$, $L_x$, $L_c$) are in units of the lattice 
constants. An external 
field $E$ drives the chains from the source near $x=1$ toward the 
substrate 
(impenetrable wall) at $x = L_x$.
The field couples with the change in energy, $E \Delta x$, where 
$\Delta x$ is the displacement of the chain node along $x$ direction.
In addition to excluded volume, there is 
a nearest-neighbor polymer-polymer repulsive and polymer-wall 
attractive 
interaction. Chains are released at a constant rate and moved with the 
Metropolis algorithm \cite{metropolis53} using segmental dynamics such as kink-jump, 
crankshaft \cite{verdier62}, or
slithering snake (reptation) or some combination \cite{binder}.  Attempt to move each chain 
node 
once is defined 
as one Monte Carlo step (MCS). The simulation is performed for a 
relatively
large number of time steps and for a sufficient number of independent samples 
to 
obtain
a reliable estimate of averaged physical quantities. 

In our earlier studies \cite{bpf, fp1, fp2}, polymer chains are 
released 
throughout the simulation.
In this study, chains are released with a constant rate for a 
sufficiently
long time (typically about three-fourths of the entire simulation time) 
before
stopping the release of new chains into the system. 
The simulation is then continued to
allow chains (already released into the system) to deposit and relax. 
As discussed below, the relaxation makes a significant difference in 
characteristics of the interface.

\section{Results}

\begin{figure}
\centerline{\includegraphics*[width=8.5cm]{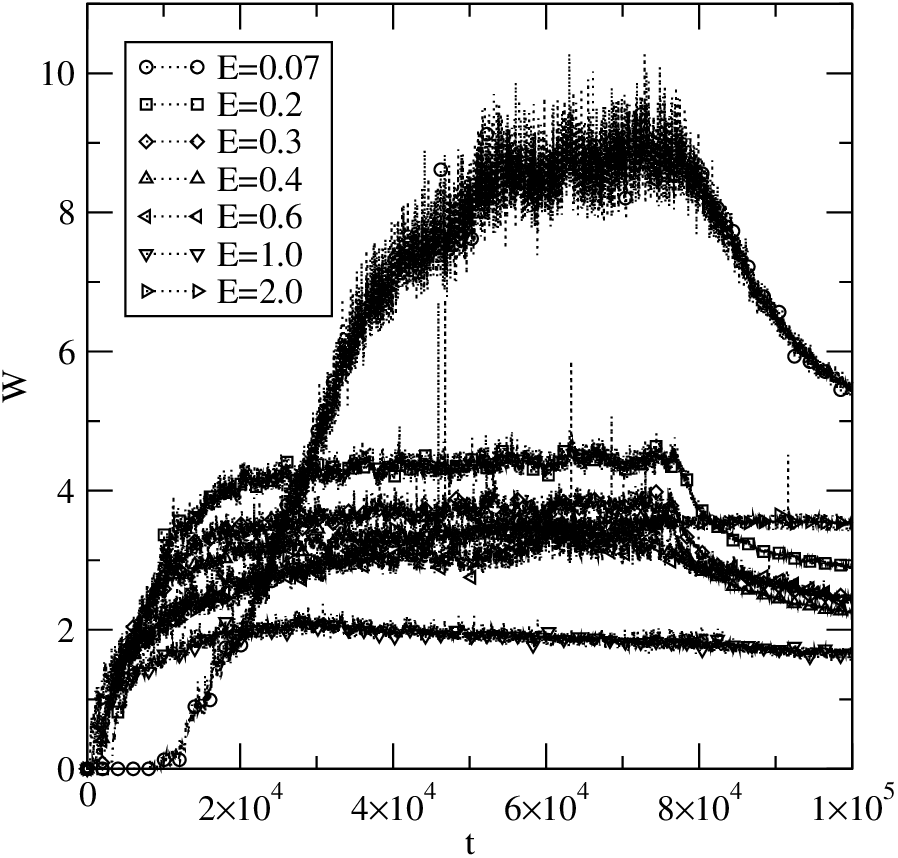}}
\caption{ 
$W$ (in units of the lattice constant) versus time steps $t$ (MCS) with 
$K$ segmental 
dynamics for different fields with 10--20 independent samples. 
}
\label{fig:w_t_k}
\end{figure}

\begin{figure}
\centerline{\includegraphics*[width=8.5cm]{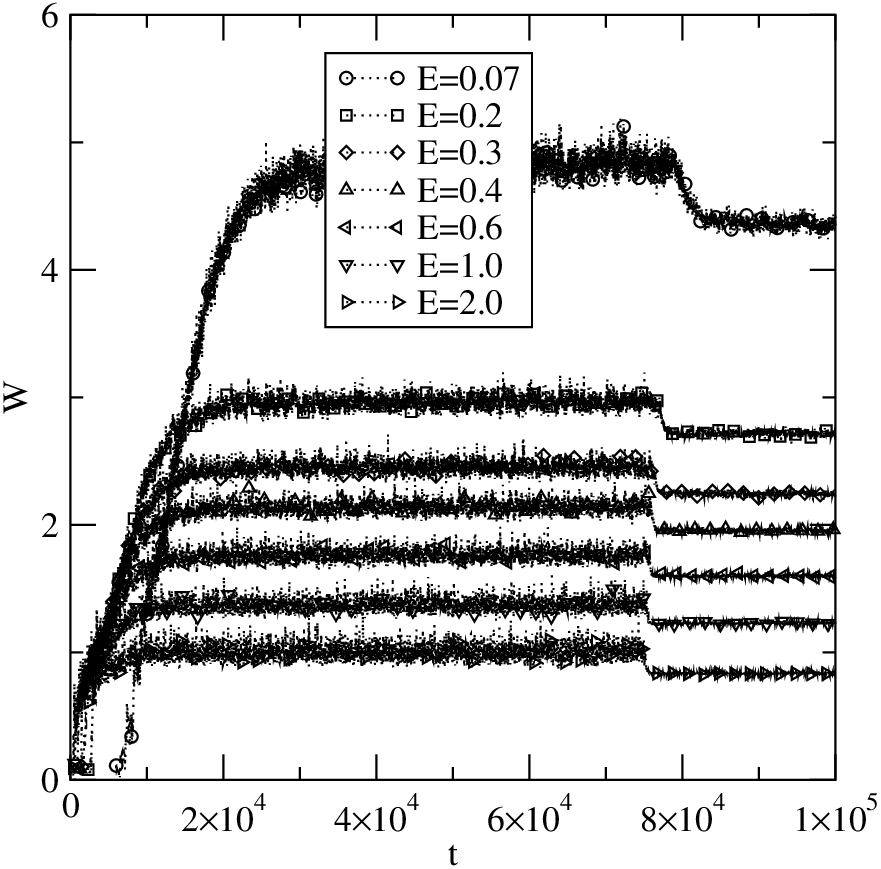}}
\caption{
$W$ (in units of the lattice constant) versus time steps $t$ (MCS) with 
$KC$ segmental 
dynamics for different fields with 10--20 independent samples. 
}
\label{fig:w_t_kc}
\end{figure}

Figures \ref{fig:w_t_k} and \ref{fig:w_t_kc} show the variation of the
interface width with the number of time steps 
for
kink-jump ($K$) (Fig.~\ref{fig:w_t_k}) and kink-jump and crankshaft ($KC$) (Fig.~\ref{fig:w_t_kc}) 
segmental dynamics for different driving fields. Data with both 
segmental 
movements are generated by ($i$) depositing chains for about 
three-fourths 
of the entire simulation steps as mentioned above and ($ii$) relaxing 
chains
already in the system (polymer bulk and the interface width) for 
the last 
one-fourth of the time steps without adding new chains.
We see a rapid growth of the interface width $W$ initially (time step 
$t \le 10^3$) particularly at higher fields before reaching a 
steady-state
value ($W_s$). At the time step, $t = 7.5 \times 10^4$, when the 
injection of
new chains stops, the interface width decays rapidly and attains a 
constant relaxed value ($W_r$). At low fields ($E = 0.07$), it takes 
longer
for the interface to both grow and decay.

\begin{figure}
\centerline{\includegraphics*[width=8.5cm]{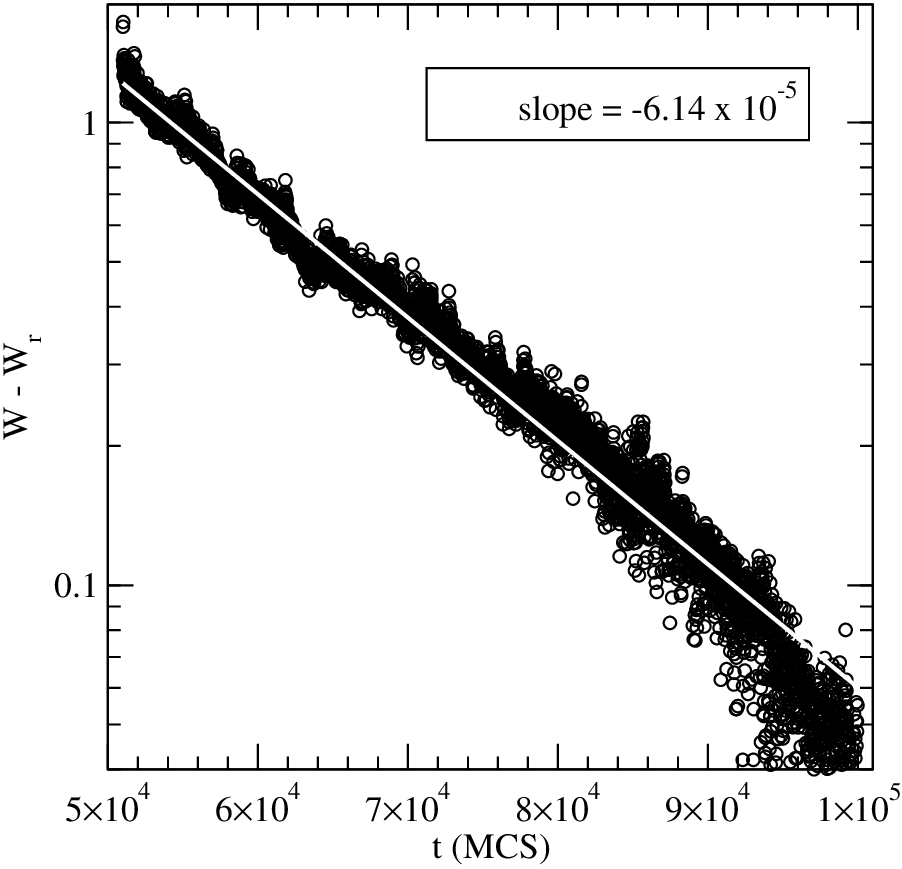}}
\caption{
Decay of $W$ (in units of the lattice constant) at $E = 0.4$, $T=1$ with 
$K$ segmental dynamics on a semilog plot.
The range of the $Y$ axis ($W-W_r$) lies between 0.4 and 1.8 and has
units of the lattice constant.
The slope of the fit is provided in the legend. Statistics
are the same as for Fig.~\ref{fig:w_t_k}.
}
\label{fig:w_t_relax}
\end{figure}

Although the general features of growth and decay of the interface width
with $K$ and $KC$ movements appear similar, there are clear differences in
the relaxation. Obviously, it is faster for the chains and the interface
width to relax with the $KC$ dynamics than with the $K$ move alone. At low
field ($E = 0.07$), the interface width is not completely relaxed
(approach a constant value) with $K$ dynamics by the end of the simulation
run. In fact, a complete relaxation even at higher values of field ($E \ge
0.6$) within the time of simulation with the $K$ dynamics is questionable.
Figure~\ref{fig:w_t_relax} shows the decay of the interface width from
steady-state to equilibrium with the $K$ dynamics during the relaxation
period. One can identify an exponential decay,
 
$$W-W_r \sim e^{-{{t-t_0}\over\tau}},$$

\noindent
where $t_0$ is the time for stopping the release of new chains, and
$\tau$ is the characteristic relaxation time for the interface
(determined by the slope). 

Initial 
interface 
growth $W \propto t^{\beta}$, ($\beta \sim 0.3$--1) has been
studied in detail \cite{bpf, fp1} and values of the growth exponent 
$\beta$ are nearly the same with both $K$ and $KC$ dynamics. It is 
clear, 
though, that the relaxation of the interface width is much faster with 
the 
$KC$ dynamics.
However, it is because of the slow relaxation of the interface width with
the $K$ dynamics that we are able to comment on the exponential decay of the 
interface width here.

Since the relaxations of chains and the interface are so different between
the $K$ and $KC$ dynamics, the major question remains, which 
dynamics 
is appropriate? In our opinion, it depends on the situation.
If one looks for the well 
behaved relaxed or steady-state properties such as the interface width,
$KC$ dynamics would be preferable over the $K$ movement of chains. 
Nevertheless, it is important to understand the details of results 
arising
from different dynamics. Lets look at the density profile of polymer
with the $K$ motion presented in Fig.~4. We immediately note the 
difference in density profile at low ($E < 0.6$) and high ($E \ge 0.6$)
fields. At low field, the polymer density remains low ($\to 0$) from 
the source end (the region for releasing new chains) and increases 
monotonically
toward the substrate. On the other hand, at high 
field there is a large density toward the source end indicating a 
build-up
(accumulation) of polymer chains before they reach the bulk region growing
from the substrate. Such a clogging restricts the deposition of polymer
chains and the growth becomes independent of the rate of polymer 
release
(i.e., dynamics-limited deposition).
Clogging occurs due to slow motion of chains with the $K$ dynamics 
alone.
Thus, care must be exercise in analyzing the interface growth, decay, 
and its scaling to field at low and high fields where deposition 
rates
differ substantially.

\begin{figure}
\centerline{\includegraphics*[width=8.5cm]{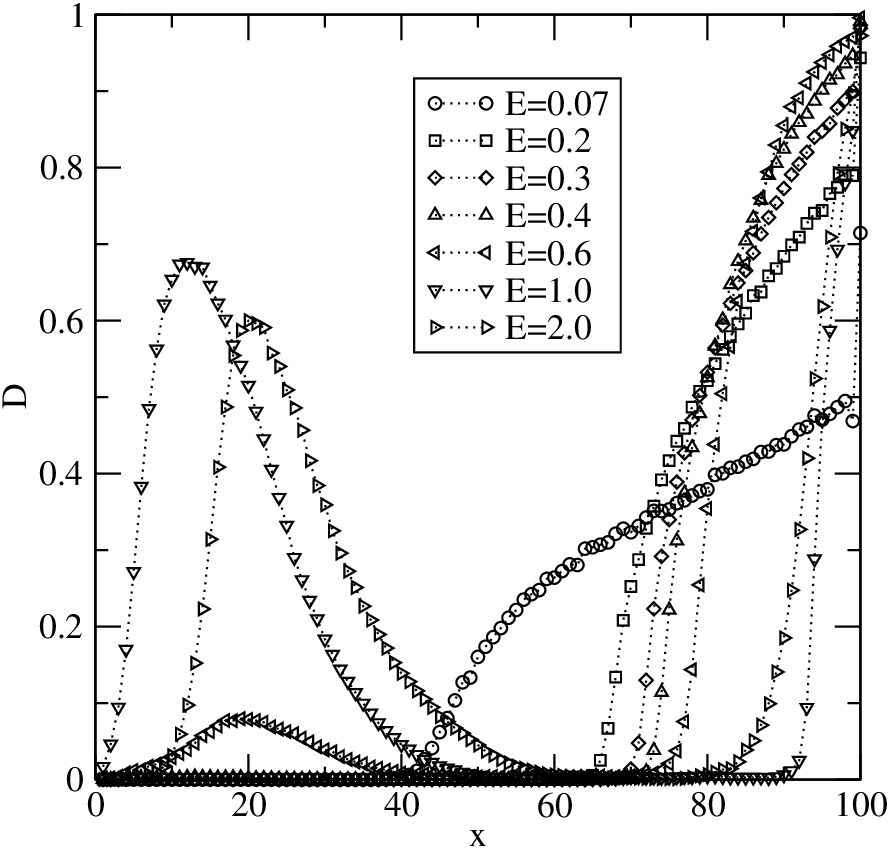}}
\caption{
Polymer density $D$ (fraction of occupied sites) vs.\ $x$ (in units of 
the lattice constant) with the $K$ 
dynamics for different
fields.
}
\end{figure}

With the $KC$ dynamics, there is no clogging at these values of field
($E \le 2.0$) and the interface width relaxes very well within our 
simulation time. Variation of the relaxed interface width $W_r$ with
the field is presented in Fig.\ 5. We see that the interface width
($W_r$) decays with the field with a power law $W_r \propto E^{-1/2}$.
Note that the nature of decay of the relaxed interface width $W_r$ 
with 
the field remains similar to that of the saturated width $W_s$ in 
steady-state. Such a decaying trend of the interface width with the 
field 
is also observed with the $K$ alone in the low field regime ($E < 
0.6$).
We have also studied the dependence of the interface width ($W_r$) on 
the molecular weight $L_c$ and find a monotonic decrease as shown in
Fig.~6. Note the contrast, while the relaxed width $W_r$ decays 
with 
the molecular weight, the saturated steady-state width $W_s$ 
increases.
In the steady-state growth, polymer chains are continuously
deposited. As a result the incoming chains at the surface are not 
relaxed.
The contribution of unrelaxed chains at the growing surface to the 
interface 
width is dominant over the relaxed chains, since the conformation of 
incoming 
chains are relatively stretched out along the direction of the field.

\begin{figure}
\centerline{\includegraphics*[width=8.5cm]{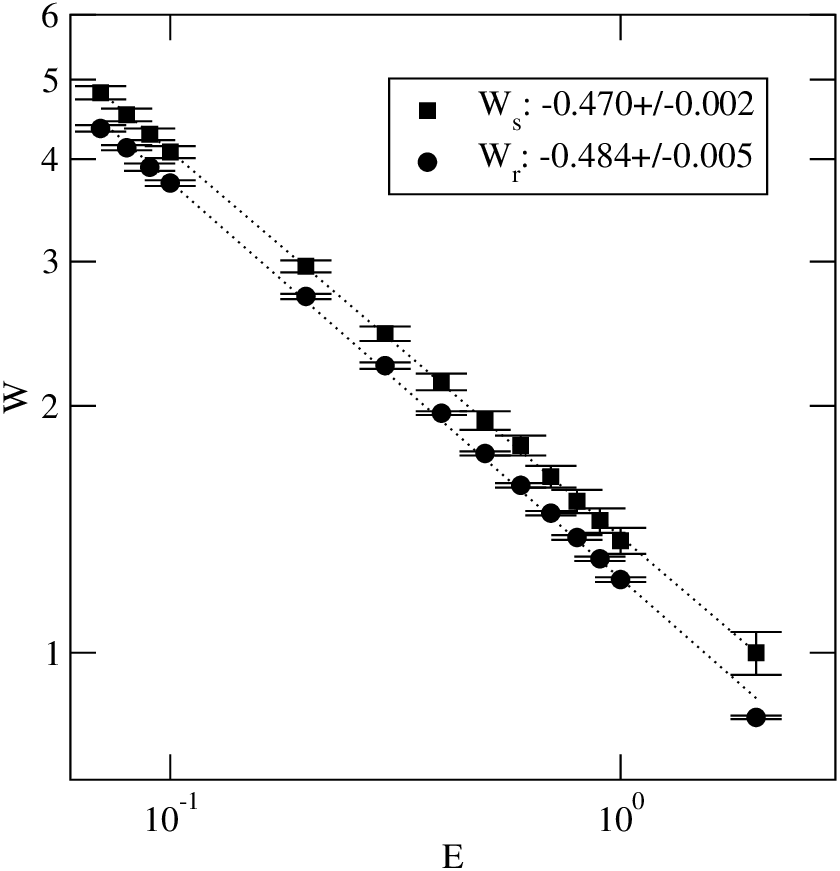}}
\caption{
Interface width ($W_s$ and $W_r$ in units of the lattice constant) vs.\ 
field 
$E$ (in arbitrary units) with 
$KC$ segmental 
dynamics. Statistics are the same as for Fig.~\ref{fig:w_t_k}.
}
\end{figure}

\begin{figure}
\centerline{\includegraphics*[width=8.5cm]{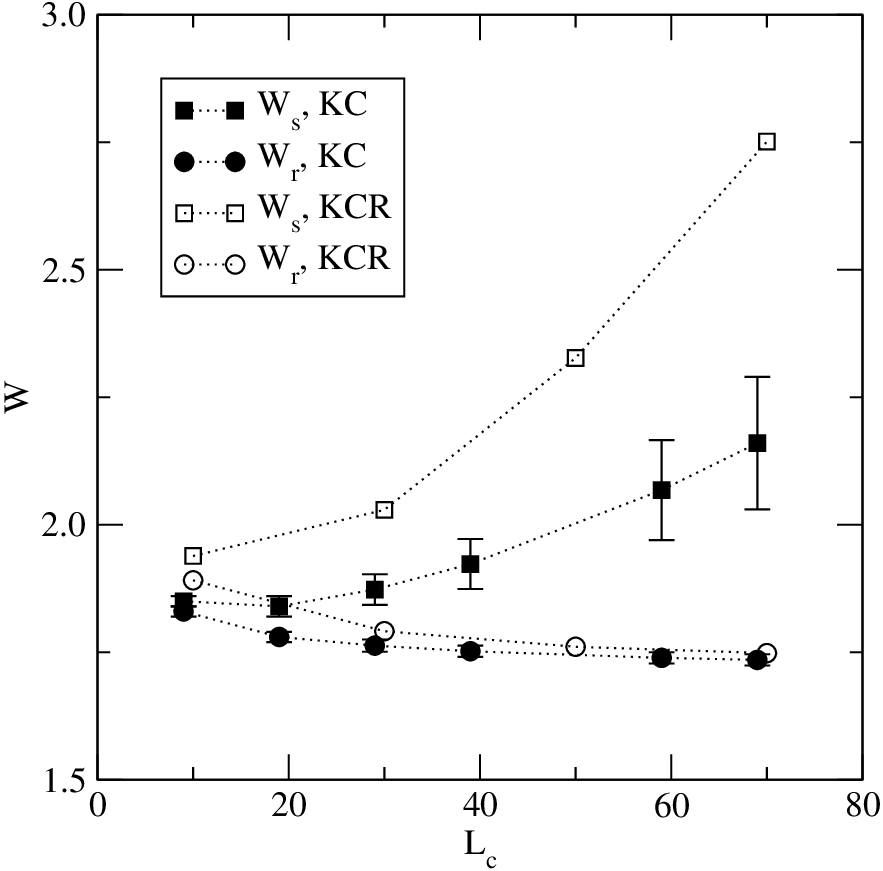}}
\caption{
Interface width ($W_s$ and $W_r$) vs.\ 
$L_c$ with $KC$ and $KCR$ segmental 
dynamics. $W_s$, $W_r$, and $L_c$ are in units of the lattice 
constant. Statistics are the same as for Fig.~\ref{fig:w_t_k}.
}
\end{figure}

We have also examined the effect of incorporating the slithering-snake
(reptation) motion to kink-jump and crankshaft dynamics ($KCR$). The
relaxation of the interface width is relatively faster with the $KCR$ than
with $KC$ and $K$ segmental movements. However, the qualitative nature of
the variation of the relaxed interface width $W_r$ with the molecular
weight $L_c$ is similar to that with $KC$ dynamics (see Fig.\ 6) with
somewhat higher magnitude of the widths. It is worth pointing out the
difference in the interface widths with $KC$ and $KCR$ segmental dynamics.
While the difference in the steady-state width $W_s$ increases with the
molecular weight, the difference in relaxed interface width $W_r$
decreases. At high molecular weights, it is rather difficult to
distinguish the relaxed interface widths. The larger difference in steady
state and relaxed interface width ($W_s-W_r$) between $KC$ and $KCR$
dynamics suggests that the magnitude of $W_s$ is dominated by the
elongation of polymer chains (along the field direction) which is larger
with the $KCR$ semental motion.  Figure 7 shows the variation of the
relaxed interface width $W_r$ with the molecular weight for two different
fields with the $KCR$ segmental dynamics.  We see that the qualitative
nature of the variation remain similar at these fields. Decay of the
interface width ($W_r$) with the molecular weight appears to be stretched
exponential. With the $KCR$ dynamics, we have also observed a power-law
decay of the interface width ($W_r$)  with the field similar to Fig.\ 5
with the $KC$ dynamics. As expected, the interface width is relatively
well relaxed with the $KC$ dynamics, and adding reptation enhances the
relaxtion further.

\begin{figure}
\centerline{\includegraphics*[width=8.5cm]{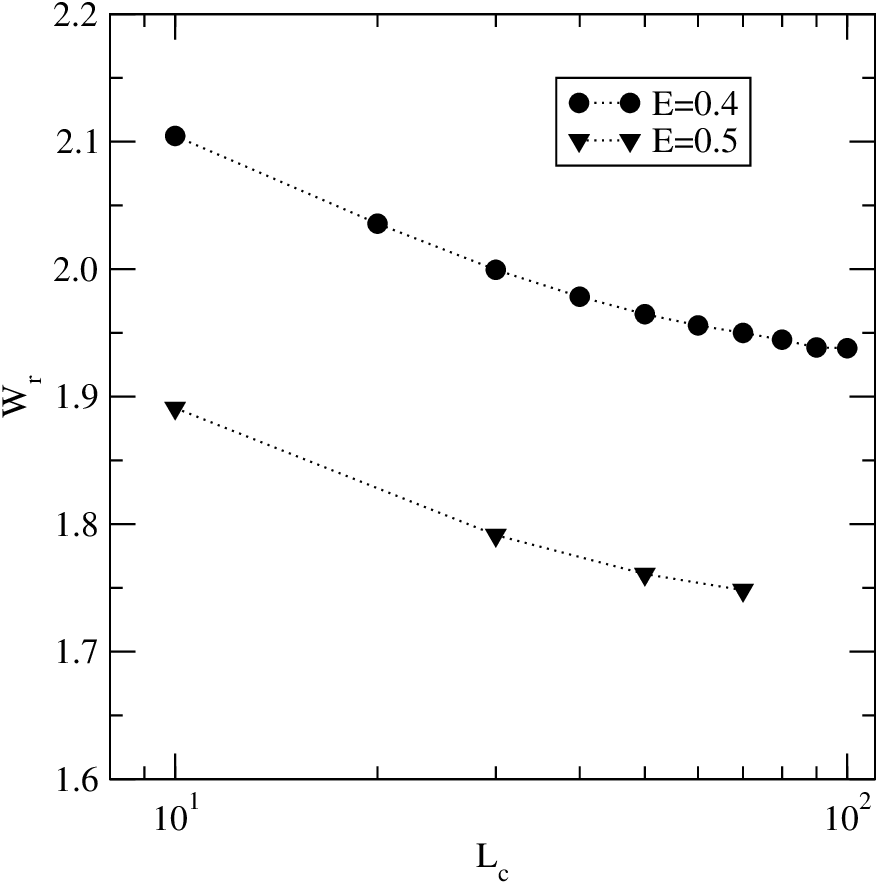}}
\caption{
Interface width ($W_r$) versus $L_c$ with 
$KCR$ segmental 
dynamics at $E=0.4$, 0.5, $T=1$. $W_r$ and $L_c$ are in units of 
the lattice constant. Statistics are the same as 
for Fig.~\ref{fig:w_t_k}.
}
\end{figure}

\section{Conclusion}

A computer simulation study was presented to investigate the effect of
segmental dynamics on the growth and decay (relaxation) of the interface width for an 
electrophoretic deposition model for polymer chains. The results for 
the 
dependence of the relaxed interface width ($W_r$) on the molecular 
weight
are quite different from that of the steady-state interface width 
($W_s$) 
reported in previous studies \cite{bpf, fp1} where polymer chains were 
continuously deposited throughout the simulation. 
In contrast to an increase of the steady-state width $W_s$, the relaxed 
width $W_r$ decays with the molecular weight.
The power-law decay of the relaxed width with the field 
($W_r \propto E^{-1/2}$) remains the same as that of the steady-state 
width ($W_s$). 

The relaxation of the width after stopping the addition of more
polymer chains to the system depends on the segmental dynamics.
For example, with the kink-jump dynamics alone, it is very difficult to
reach relaxed interface width within a reasonable simulation time at 
low
fields ($E \le 0.07$). On the other hand, clogging occurs around the 
entrance
area of polymer injection at high fields ($E \ge 0.6$), which reduces 
the rate
of polymer deposition on the substrate. Thus, the scaling of the 
interface
width at high fields should be different from those at low to moderate
field values. Inclusion of crankshaft motion leads to faster 
growth
and interface relaxation at all field values we studied. Adding 
large scale
segmental dynamics (reptation) enhances the interface relaxation and 
reduces
the magnitude of the relaxed interface width. The scaling of the 
interface 
width with the field and molecular weight is qualitative similar. 
 
\acknowledgments
 
We acknowledge partial support from a DOE-EPSCoR grant.

\end{document}